\documentclass{mn2e}

\usepackage[dvips]{graphicx}

\begin{document}

\title[Baryons and dark matter haloes]{The impact of baryons on dark matter haloes}

\author[Pedrosa, Tissera, Scannapieco]{Susana Pedrosa$^{1,2}$, Patricia B. Tissera$^{1,2}$ and Cecilia Scannapieco$^{3}$\\
$^1$  Consejo Nacional de Investigaciones Cient\'{\i}ficas y T\'ecnicas, CONICET, Argentina.\\
$^2$ Instituto de Astronom\'{\i}a y F\'{\i}sica del Espacio, Casilla de Correos 67, Suc. 28, 1428, Buenos Aires, Argentina.\\
$^3$ Max-Planck Institute for Astrophysics, Karl-Schwarzchild Str. 1, D85748, Garching, Germany\\
}

\maketitle

\begin{abstract}

We analyse the dark matter (DM) distribution in a  $\approx 10^{12} M_{\sun}$ halo extracted from a simulation consistent with the concordance cosmology, where the physics regulating the transformation of gas into stars was allowed to change producing 
galaxies with different morphologies. Although the DM profiles get more concentrated as baryons are collected at the centre of the haloes compared to a pure dynamical run, the total baryonic mass alone is not enough to fully predict the
reaction of the DM profile. We also note that baryons affect the DM distribution even outside the central regions. Those systems where the transformation of gas into stars is regulated by Supernova (SN) feedback, so that significant disc structures are able to form, are found to have more concentrated dark matter profiles than a galaxy  which has efficiently transformed most of its baryons into stars at early times. The accretion of satellites  is found to be
associated with an expansion of the dark matter profiles, triggered by angular momentum transfer from the incoming satellites. As the impact of SN feedback increases, the satellites get less massive and are even strongly disrupted before getting close to the main structure causing less angular momentum transfer. Our findings suggest that the response of the DM halo is driven by the history of  assembly of baryons into a  galaxy along their merger tree.

\end{abstract}

\section{Introduction}

Dark matter haloes in Cold Dark Matter (CDM) scenarios have been extensively studied during the last
decades (e.g. Merrit et al. 2006; Prada et al. 2006; Gao et al. 2008; Navarro et al. 2008), since understanding their properties is relevant not only for the study of galaxy formation but also for
constraining the nature of DM itself (e.g. Springel et al. 2008). Several controversies have been issues of thoughtful discussions such as the universality of the DM profiles (Navarro, Frenk \& White 1996, NFW)  
or the systematically increasing central DM density with  increasing numerical resolution  (Moore et al. 1999).
However, new improvements in the numerical codes and computational power have challenged previous results.
Recently,  Navarro et al. (2008) found evidence for the
non-universality of the DM profiles  and for non-cuspy inner profiles by using high resolution N-body  simulations.

The assembly of  baryons in galaxies is expected to drive changes in the DM distribution 
 as it reacts to gas infall and outflows (e.g. Read \& Gilmore 2005) and to the angular
momentum redistribution caused by infalling satellites (Barnes \& Hernquist 1996; Debattista et al. 2008, D08).
 Blumenthal et al. (1986, B86; see also Young 1980 for a different implementation) proposed the adiabatic contraction (AC) model to globally predict the effects 
of central baryonic accumulation on the DM distribution. Although it is clear that the condensation of baryons increases the DM
concentration, it remains to be determined how this would proceed and how  the transformation
of baryons into stars and supernovae (SN) feedback might affect the DM distribution.
Gnedin \& Zhao (2002) studied the effects of maximum feedback on the central density of DM haloes of gas-rich 
dwarfs, finding them to be too weak to explain
 the observed rotation curves. Interestingly, Read \& Gilmore (2005) showed that two impulsive mass-loss phases
 are needed to significantly change the DM profiles of dwarf galaxies, suggesting the need for bursty star formation histories.
Also, several works have studied the change in shape of DM haloes toward rounder configurations induced by the presence of a central baryonic concentration (Kazantsidis et al. 2004).  D08 claimed that a larger, irreversible change in the DM distribution can 
be produced if there is  transport of angular momentum from baryons to the DM.
In a fully cosmological context,  simulations  have already shown how the DM haloes concentrate with increasing baryonic condensation, providing hints of a possible dependence on the assembly history of baryons
(e.g. Tissera \& Dominguez-Tenreiro 1998; Gnedin et al. 2004; O\~norbe et al. 2007). However, the high complexity of the problem requires high numerical resolution to describe in more detail the joint evolution of  the DM and 
the baryonic distributions to get to more robust results. Also, none of these previous studies included treatments for
 SN feedback and a multiphase interstellar medium (ISM) self-consistently. This  is of particular
relevance since feedback not only regulates the star formation in the main galaxy, but also in the  merging substructures and, as a consequence, it might also change the history of angular momentum transfer process which plays a major role in  hierarchical scenarios (e.g. Navarro et al. 2004; D08).

We study  a set of intermediate-resolution  cosmological simulations where  different baryonic structures
 have been formed from the same initial condition (IC), through the modification of the physics of baryons
(Scannapieco et al. 2008, S08). These experiments are taken as toy models to analyse how the dark matter evolves when baryons are assembled in a different fashion.
In this letter, we focus our discussion on  four experiments which better illustrate the main results (for further details see Pedrosa et al. in preparation).

\section{Simulations and Analysis}

We analysed four simulations of a  $\approx 10^{12} M_{\sun}$ type galaxy run with an
extended version of the code {\small GADGET-2}  (Scannapieco et al. 2005, 2006). Although all cosmological numerical simulations to date suffer from a number of limitations such as coarse-resolved star formation and interstellar medium physics, this extended {\small GADGET-2} code has been designed to improve the respresentation of the ISM and SN feedback (S08).
Our simulations are fully cosmological and include a multiphase model for the gas component, metal-dependent
 cooling, chemical enrichment and energy feedback by type II and Ia SN. The IC corresponds to a $\approx 10^{12} $M$_{\odot}$ halo extracted from a cosmological simulation and re-simulated with higher resolution. This halo was required to have no major mergers since $z=1$. 
The simulations are consistent with  a $\Lambda$CDM Universe with $\Omega_{\Lambda}=0.7$, $\Omega_{m}=0.3$, $\Omega_{b}=0.04$, 
$\sigma_{8}=0.9$ and $H_{0}= 100 h \ {\rm km} \ {\rm s}^{-1}\ {\rm Mpc}^{-1}$, with $h=0.7$. The particle mass 
is $1.6\times 10^{7} h^{-1}\ $M$_{\odot}$ and $2.4\times 10^{6} h^{-1}\  $M$_{\odot}$ for the dark 
matter and baryonic particles, respectively. 
 The maximum gravitational softening used was $\epsilon_{\rm g}=0.8 h^{-1}$ kpc. At $z=0$, the haloes are  relaxed as indicated by a relaxation parameter of $\approx 0.002$ (Neto et al. 2007).
All analysed DM  haloes have more than 120000 particles within their virial radius (the total number of particles goes from $\approx 170000$ to $\approx 320000$ depending on the star formation histories).
We analysed three simulations (NF, E-0.7 and E-3) 
of the same IC but using  different input
parameters for the SN feedback model. 
As a result, DM haloes finally host baryonic structures with different morphologies.
In particular, simulation NF has been run without including SN feedback, while
E-0.7 and E-3  include feedback assuming an energy per SN of $0.7 \times 10^{51}$ and $3 \times 10^{51}$ erg, respectively. For comparison, we also performed a pure gravitational run (DM-only) of the same IC (particle mass of $1.84\times 10^{7} h^{-1}\ $M$_{\odot}$).

A full discussion of the effects of SN feedback on the 
baryonic dynamics and star formation histories can be found in S08.
Here, we only summarize the main characteristics of the galaxies 
which are relevant to our discussion. 
At $z=0$, the galaxy formed in  NF is dominated by a extended spheroid, with most of its stars
formed at $ z > 2$.
E-0.7 has been able to produce a galaxy with an important disc component
as a result of the regulation of star formation by SN feedback. This system has  a half mass radius of 5.72 $h^{-1}$ kpc  and a mass disc-to-spheroid ratio of 0.82.
In E-3, the large amount of energy assumed per SN triggers  violent 
outflows which expel a significant amount of the gas content of the main galaxy, producing a small thick stellar disc.
Table \ref{tab1} lists the final stellar masses of
the galaxies formed in these simulations.

\begin{table}
  \begin{center}
  \caption{Main characteristics of  DM haloes and the main central galaxy. 
 We show the total stellar mass $M_{\rm s}$ of the central galaxy,
the  total-mass-to-stellar
mass ratio $M_{\rm t}/M_{\rm s}$, the $n$ and $r_{-2}$ Einasto parameters and the
 inner logarithm slope 
$\gamma_{\rm inner}$. All masses are evaluated at two times the radius
that enclosed $83\%$ of the baryons in the central regions. Bootstrap errors for $n$ 
and $r_{-2}$ are shown within parenthesis.}
  \label{tab1}
 {\scriptsize
  \begin{tabular}{|l|c|c|c|c|c|}\hline
{\bf Run}  & {\bf $M_{\rm s}$} &  {\bf $M_{\rm t}/M_{\rm s}$} &  {\bf $n$} & {\bf $r_{-2}$} & {\bf $\gamma_{\rm inner}$} \\
   &$10^{10} h^{-1}\ $M$_{\odot}$& &  & kpc $h^{-1}$&  \\ \hline
NF &  15.9 &  4.1 &  5.973 (1)& 16.90 (1) & 1.24  \\ 
E-0.7 &  7.5 &  5.1 &  6.887 (4)& 15.36 (1)& 1.28  \\ 
E-3 &  1.3 & 37.2 & 5.585 (1)& 21.89 (1)& 1.15  \\ 
DM-only &  - & - & 5.239 (3) & 24.06 (1) & 1.08 \\ 
DNF & 0.018 &  2.4  & 7.164 (5)& 1.10 (1) & 1.30  \\ 
DE-0.7 &  1e-4 &  1486 & 5.445 (4)& 2.17 (1)& 1.00\\
D-only &  - & - & 5.222 (3) & 2.41 (1) & 1.00 \\ 
  \end{tabular}
  }
 \end{center}
\vspace{0.1mm}
\end{table}

We constructed the DM profiles for the simulated galaxies at $z=0$,
after cleaning them from substructures.
 We fitted  the NFW, Jaffe (1983) and Einasto (1965) expressions to the DM profiles, between three times the gravitational
 softening and the virial radius, finding that the Einasto model provides the best fit
in all cases.
 The Einasto formula has  three free parameters, $n$, $r_{-2}$  and $\rho_{-2}$,
 which indicate the sharpness of the profiles, and the radius and density where their logarithmic slope takes the isothermal value. 
 The fitting values obtained for the different simulations are shown in Table \ref{tab1}. 
We estimated bootstrap errors for $n$ and $r_{-2}$ by fitting the Einasto's formula to 100 
randomly-generated realizations
of the DM  profiles and by estimating the standard dispersion over the generated set of  parameters. 

\begin{figure}
\resizebox{8cm}{!}{\includegraphics{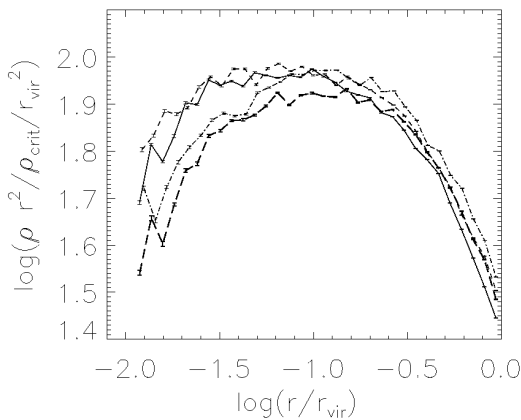}}
\hspace*{-0.2cm}
\caption{Dark matter profiles   for the NF (solid line), E-0.7 (short-dashed line), E-3 (dotted-dashed line) and  DM-only
 (thick long-dashed line) experiments. The inner bin corresponds to three gravitational softenings.
The errorbars have been estimated by boostrap resampling technique. } 
\label{fig1}
\end{figure}

When baryons are considered, the shape of the DM profiles in the 
central regions changes significantly in comparison to the DM-only
case, as it is shown in  Fig.~\ref{fig1}. 
The profile of the DM-only run is sharper (i.e. smaller $n$ value)
than those of haloes with baryons, indicating the increase in the DM concentration in the latter cases.
In E-3, where an extreme (and perhaps unrealistic) value of $3\times10^{51}$ erg was assumed, we obtained the less concentrated  DM profile among the cases with baryons. In fact its
DM profile is weakly more concentrated than the DM-only one. 
As expected, the galaxy in E-3 is also the most DM dominated one in the central region, as indicated by the total
to stellar mass ratio $M_{\rm t}/M_{\rm s}$ (Table~\ref{tab1}). 
In this simulation, most of the gas has been blown away and  only a
 small fraction of stars has been formed in a bursty fashion.

\begin{figure}
\hspace*{-0.2cm}\resizebox{7cm}{!}{\includegraphics{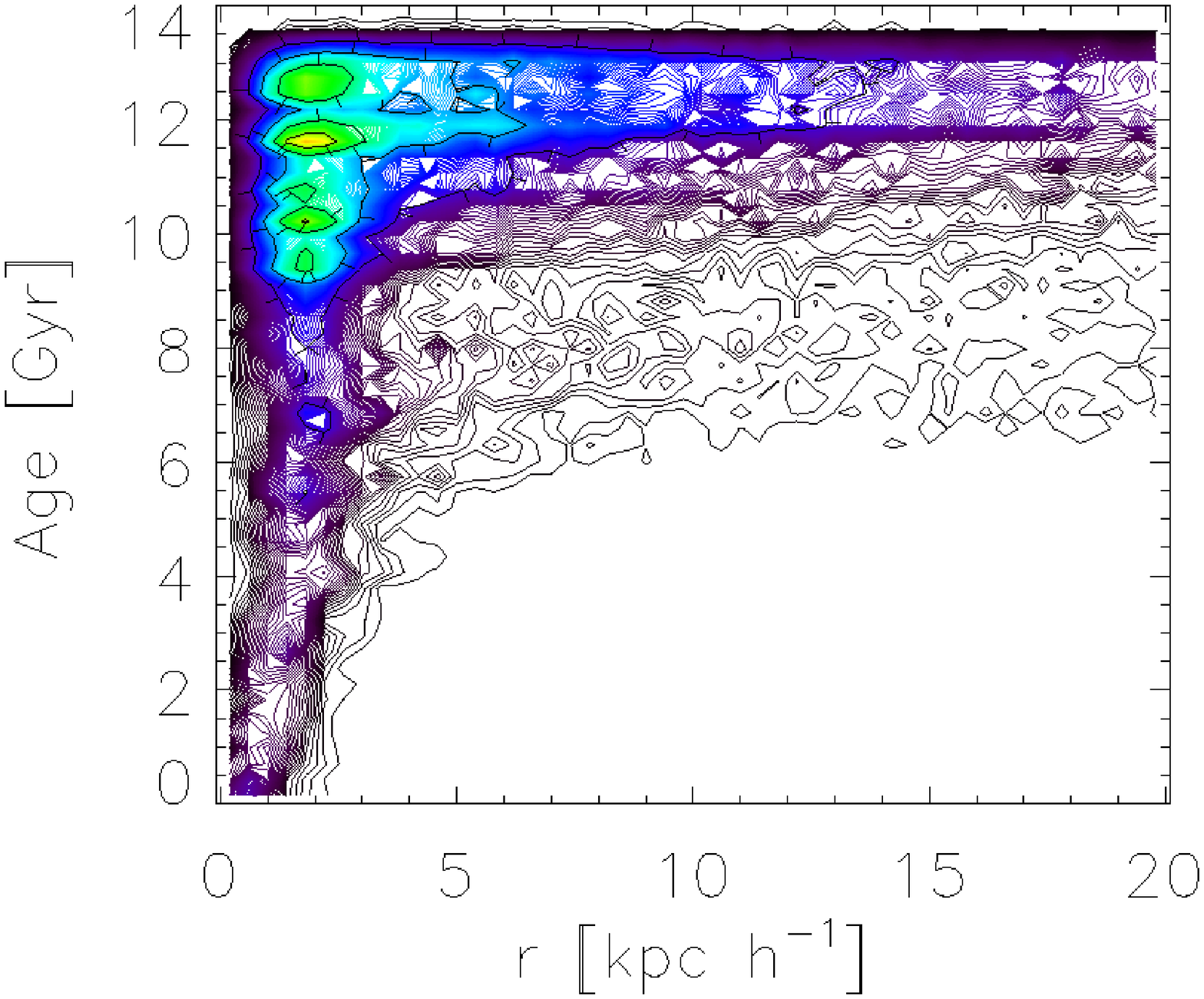}}
\hspace*{-0.2cm}\resizebox{7cm}{!}{\includegraphics{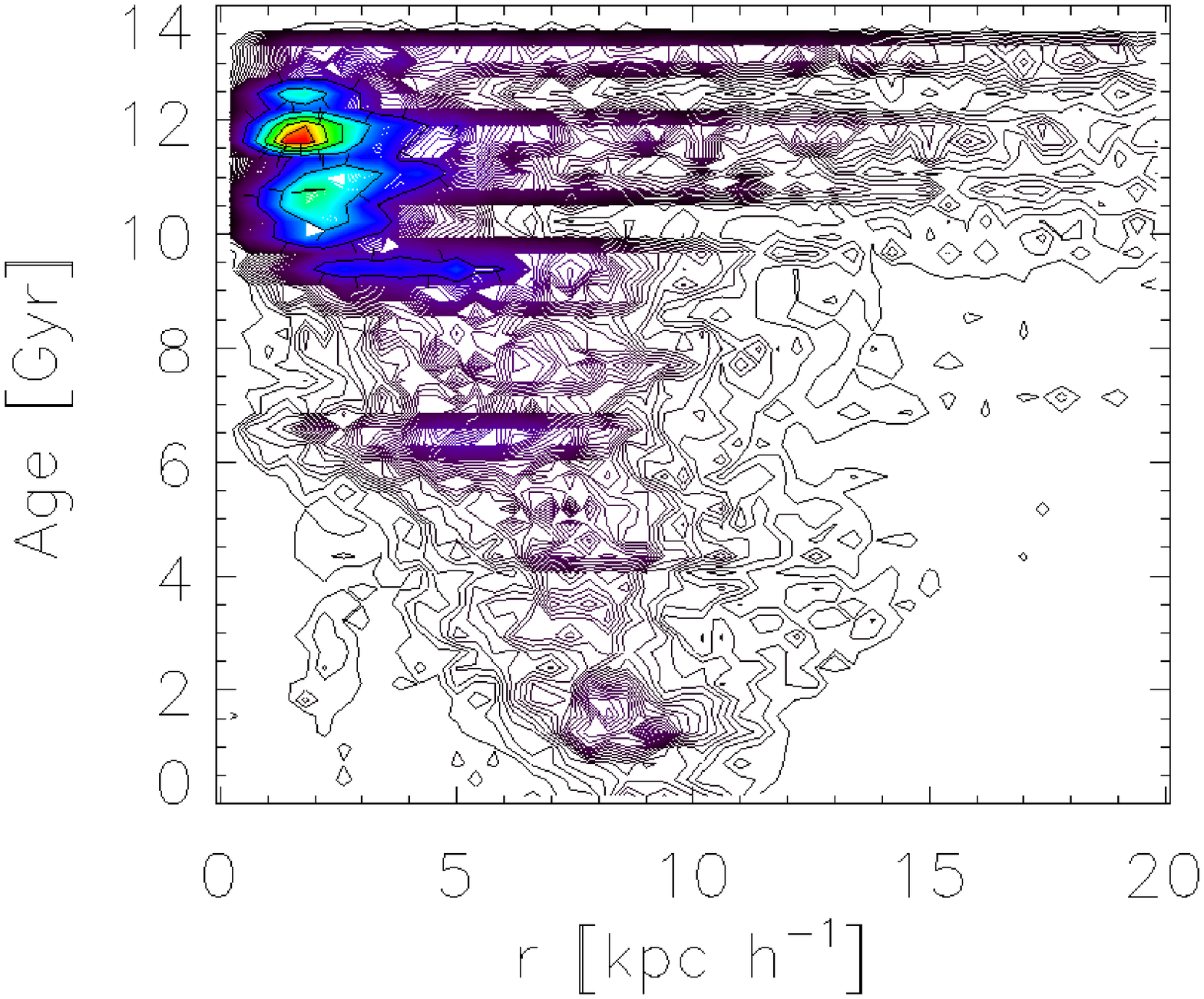}}
\hspace*{-0.2cm}\\
\caption{Age-radial distance maps of the stars in the NF (upper) and E-0.7 (lower) experiments. This figure shows the inside-out formation of the disc in E-0.7 and the outside-in formation of the spheroid in NF.} 
\label{histories}
\end{figure}

The comparison between the  E-0.7 and the NF profiles shows that the DM distribution in E-0.7  is slightly more concentrated than in NF, although it hosts a galaxy a factor of two less massive
than the later.  This finding suggests that the total amount of baryons collected within the central region of a halo is  not the only relevant factor  affecting the response of the DM to the presence of baryons (this behaviour is common to the full S08 simulation sample as it will be discussed in Pedrosa et al. in preparation). The NF and  E-0.7 runs have produced very different galaxies as we have already explained. In Fig.\ref{histories} we show their  age-radial position  maps where the
star formation histories and spatial distribution of the stars  can be clearly appreciated. The galaxy in the  NF run is dominated by
old stars, determining an extended spheroid with  $78\%$ of the final stellar mass older than 10 Gyr. The system in  E-0.7 has a compact old spheroid and an important disc component populated by younger stars. This disc survives the interaction with satellites at lower redshifts (S08). The different star formation histories and morphologies in NF and E-07 are the result of the action of the feedback which, in the E-0.7 regulates the transformation of gas into stars preventing the formation of an early, extended spheroid and assuring the existence of gas to form a disc later on. The SN feedback has also affected the formation of stars in the satellite systems accreted by the main system, so that in the NF run they are more massive.

\begin{figure}
\resizebox{8cm}{!}{\includegraphics{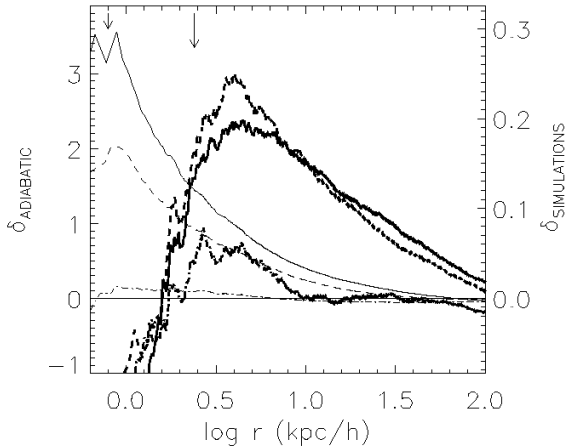}}
\hspace*{-0.2cm}
\caption{ Residuals from the DM circular velocity measured in the simulations (thick lines) and from the AC prescription (thin lines) applied to the pure dynamical run considering the baryonic distribution in the NF (solid lines), E-07 (dashed lines) and
the E-3 (dotted-dashed lines) simulations. The arrows denotes one (short) and three (long) gravitational softenings}.
\label{residuos_letter}
\end{figure}

We compare the residuals of the circular velocities estimated by applying the AC method of B86 to the DM-only run taking into account the baryonic distributions of the corresponding hydrodynamical runs (thin lines) and those obtained directly from the simulations (thick lines) as shown in Fig. ~\ref{residuos_letter}. 
All residuals are estimated with respect to the circular velocity of the DM-only run. 
We found that the B86 model not only overpredicts the level of concentration but also changes the shape of the DM distribution when compared to the DM profiles obtained from the cosmological runs. While in the B86 method, the effects are principally determined by the amount of baryons accumulated at the centre, the simulated profiles are the result of the joint evolution of baryons and DM. More sophisticated AC methods could account for these differences as suggested by the results of  Gnedin et al. (2004) and  Sellwood \& McGaugh (2005). We will explore this possibility in a forthcoming paper.

As can be seen in Fig. ~\ref{residuos_letter}, while the simulations yield  the largest residuals for the E-0.7 run which has the important disc component, the B86 model predicts the strongest effect for the NF run because it has the largest baryonic mass system,  almost twice more massive than in E07. In the case of E-3 similar level of residuals are obtained. In the very central regions (within one $\ \epsilon_{\rm g}$  and three $\ \epsilon_{\rm g} $) our simulations predict an expansion of the DM profiles when baryons are present which can be related to the evolution of the systems as discussed below. We also found this behaviour in the higher numerical resolution haloes of the Aquarius projects (Springel et al. 2008b).

\begin{figure}
\resizebox{8cm}{!}{\includegraphics{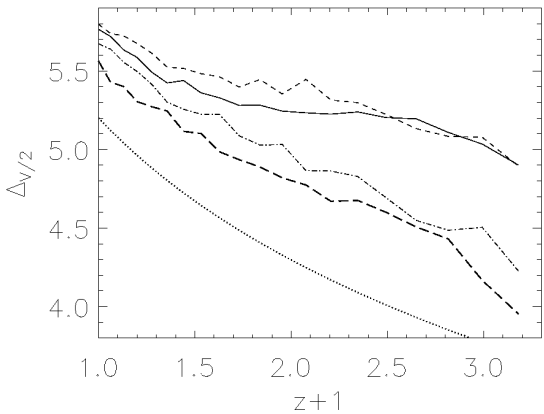}}
\hspace*{-0.2cm}
\caption{ Central halo mass concentration $\Delta_{v/2}$ as function of redshift 
 for the NF (solid line), E-0.7 (short-dashed line), E-0.3 (dotted-dashed line) and  DM-only (long-dashed line) experiments. The solid line is the expected growth due to the expansion of the Universe.} 
\label{fig2}
\end{figure}

These results suggest a connection between the DM evolution and the history of formation of the baryonic structures. To study this point, we estimated the concentration parameter proposed by Alam, Bullock \& Weinberg (2002), $\Delta_{v/2}$\footnote{This parameter measures the mean DM density normalized to the cosmic closure density within the radius 
at which the circular rotational speed due to the DM alone rises to half its maximum value.}.
 This parameter has the advantage of being independent of a specific density profile and, as it is an integrated quantity, it 
 can be estimated more robustly at any time.
In  Fig.~\ref{fig2} we show $\Delta_{v/2}$  as a function of redshift for our set of simulations.
All haloes increase their central concentration as they grow with time. The solid line shows the expected
relation (i.e. hereafter critical relation) due to the expansion of the Universe alone.
The DM-only run has the lowest DM concentration, as expected, at all times, although
it flattens between $z \approx 1$ and $z \approx 1.8$, coinciding with the close approaching
of two satellites.
The transfer of angular momentum from the satellites to the dark matter can explain its subsequent
expansion (Pedrosa et  al. in preparation).
In E-3, $\Delta_{v/2}$ grows following the slope of the critical relation even closer than in the DM-only run since the entrances of satellites cause weaker effects on the DM distribution because they are less massive due to the strong action of SN feedback.
An interesting behaviour is determined by the NF and E-0.7 runs.
 Both haloes have similar concentrations until $z \approx 1.5$, and from there on, 
the  E-0.7 halo  has systematically higher values of  $\Delta_{v/2}$ than the NF one. 
The analysis of the baryonic dynamics of these tests shows 
that by this redshift the disc component in E-0.7  starts forming.
The incoming satellites are less massive and the gas is able to settle onto a disc smoothly. This and 
the fact that the stellar disc is stable produce a smaller effective transfer of angular momentum to
the dark matter which, at the end, is slightly more concentrated than in the NF. 

The shape in the central region of the DM profiles can be also quantified by calculating the logarithm 
slope $\gamma(r)= -$dln$\rho /$dln $r$ at the innermost resolved radius. 
We found $\gamma_{\rm inner} \approx 1$ for the DM-only run (Table~\ref{tab1}), in agreement with recent estimations of Navarro et al. (2008)  for  galaxy-sized DM haloes in  the Aquarius Project. The presence of baryons increases 
$\gamma_{\rm inner}$ to $ \approx 1.25 $ as it can be seen in  Table~\ref{tab1}.

The shape of the dark matter profiles in dwarf galaxies is one of the main caveats of the CDM cosmology.
And it is also a difficult problem to tackle by numerical codes due to limited numerical
resolution and oversimplified modelling of baryonic physics. To shed light on this point, we studied  a scaled-down version of our $\approx 10^{12}$M$_{\odot}$ type galaxy designed to get a system
of $10^{9} $M$_{\odot}$. The maximum physical softening used in this run is  80 pc h$^{-1}$. This experiment  (DE-0.7) has been run  with the same cosmic and feedback parameters as  E-0.7. However, the  effects of the  SN feedback on baryons are much stronger because of
its shallower potential well as explained in detail by S08.
We compared the DM profiles of this DE-0.7 run with those obtained in a no feedback (DNF) case and in
a pure dynamical run (D-only) of the same scaled-down galaxy.

\begin{figure}
\resizebox{8cm}{!}{\includegraphics{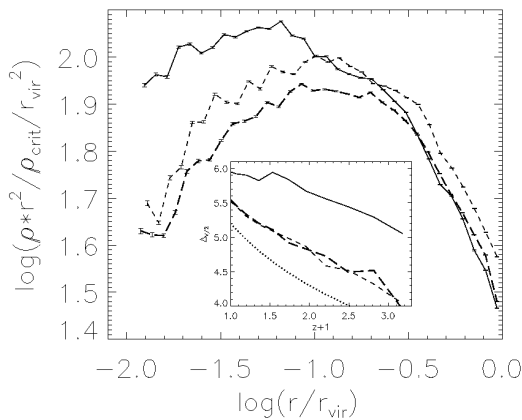}}
\hspace*{-0.1cm}
\caption{ Dark matter density profiles for the scaled-down system of  $10^{9} h^{-1}\ $M$_{\odot}$ virial mass,
for the DNF (solid lines), DE-0.7 (short-dashed lines) and the D-only (long-dashed lines) experiments. 
The inner bin corresponds to three gravitational softenings. The inset plot shows the corresponding
 $\Delta_{v/2}$ concentration parameters as a function of redshift.}
\label{fig3}
\end{figure}

 For this small galaxy, SN feedback strongly prevents  the star formation 
activity compared with that of the DNF case. In this shallow potential well, violent 
winds  are able to blow out an important fraction of  the baryons, 
inducing the system to form stars in a series of very weak starbursts after an early important one.
 As a consequence the DM profile is more similar to that of the pure dynamical run (Fig.~\ref{fig3}).
 The inner logarithm slope of the DM profile in the DE-07 and DNF are $\gamma_{\rm inner} = 1.0 $ and $\gamma_{\rm inner} = 1.3$, respectively. Conversely in the DNF case the central accumulation of baryons induced a higher concentration of DM compared to the D-only run. As shown in Table 1, the shape parameter of the Einasto model goes from $n=5.2$ in the D-only to $n=5.4$ and $n=7.2$ in the DE-07 and DNF, respectively. We found that the evolution of $\Delta_{v/2}$ with time of the DM profile in the DE-0.7 is very similar to that of the pure dynamical run in agreement with
previous results (Read, Pontzen and Viel 2006). This could be caused by the too strong SN feedback effect which blows away most of the baryons or by deficient numerical resolution. Indeed, by using very high
numerical resolution simulations and considering molecular cooling, Mashchenko, Wadsley and Couchmanet (2008) detected strong differences in the DM profiles at $z=5$.

\section{Conclusions}

We found that, in all our $10^{12}$ haloes, including the baryons resulted in a more concentrated DM density distribution compared to the equivalent pure DM simulation. However, using the total baryonic mass that collects at the centre as input to the AC model of B86, we were not able to reproduce our final DM density distributions. 
Our results indicate that the way baryons are assembled into a galaxy affect the DM evolution so that
its final distribution also depends on the morphology of the baryonic system. 
An efficient star formation activity produces satellites with larger stellar masses which are able to survive further in the halo. This could cause a larger angular momentum transfer (D08) which, on its turn, might explain the expansion of the DM central concentration. The treatment of SN feedback prevents this early overproduction of stars. 
Hence the gas is allowed to get to the centre before being transformed into stars forming a more compact stellar spheroid at early times.
 At later times, the remanent gas and the disrupted satellites are  able to settle onto more important disc structures.
When SN feedback is very efficient and blows away significant masses of gas, then the DM profile is less affected because of both the low central baryonic concentration and the easier dissolution of smaller satellites that get into the halo. We found that this behaviour can be linked to a larger DM concentration presumedly due to a weaker angular momentum transfer from the satellites to the dark matter. 
Our findings confirm previous lower resolution results and suggest new clues for the understanding of galaxy formation, which will be explored in a forthcoming paper.

\section*{acknowledgement}
We thank the anonymous referee for his/her valuable comments. This work was partially supported by PICT 32342(2005), PICT  Max Planck 245(2006) of Foncyt and DAAD-Mincyt collaboration(2007).


\begin{thebibliography}{}

\bibitem[Blumenthal,G.R. et al. (1986)]{Blumen86}
{Blumenthal,G.R., Faber,S.M., Flores,R. \& Primack,J.R.} 1986,
\textit{ApJ}, 301, 27

\bibitem[Bullock, J. et al. (2001)]{Bull01}
{Bullock, J.S., Kolatt, T.S., Sigad, Y., Somerville, R.S., Kravtsov, A.V., Klypin, A.A., Primack, J.R. \& Dekel, A.} 2001,
\textit{MNRAS}, 321, 559

\bibitem[Choi, J.H. et al. (2006)]{Choi06}
{Choi,J.H., Lu,Y., Mo,H.J. \& Weinberg,M.D.} 2006,
\textit{MNRAS}, 372, 1869

\bibitem[Debattista, V. P. et al. (2008)]{D08}
{Debattista, V.P., Moore, B., Quinn, T., Kazantzidis, S., Maas, R., Mayer, L., Read, J., Stadel, J.} 2008,
\textit{ApJ}, 681, 1076

\bibitem[Navarro, J. F. et al. (1997)]{Nav97}
{Navarro J.F., Frenk C.S. \& White S.} 1997,
\textit{ApJ}, 490, 493

\bibitem[Dutton, A.A. et al. (2008)]{Dutton08}
{Dutton, A.A., van den Bosch, F.C. \& Courteau, S.} 2008,
\textit{arXiv0801.1505}

\bibitem[Einasto, J. (1965)]{Einas65}
{Einasto J.} 1965,
\textit{Trudy Inst. Astrofiz. Alma-Ata}, 51, 87

\bibitem[Gao et al. (2008)]{Gao08}
{Gao,L., Navarro,J.F., Cole,S., Frenk,C., White,S., Springel,V., Jenkins,A., Neto, A.F.} 2008,
\textit{MNRAS}, 387, 536

\bibitem[Gnedin, O.Y. \& Zhao, H. (2002)]{Gnedin02}
{Gnedin,O.Y. \& Zhao,H.} 2002,
\textit{MNRAS}, 333, 299

\bibitem[Gnedin, O.Y., Andrey (2004)]{Gnedin04}
{Gnedin,O.Y. Kravtsov, A.V., Klypin, A.A. \& Nagai D.} 2004,
\textit{ApJ}, 616, 16

\bibitem[Jaffe, W. (1983)]{Jafffe83}
{Jaffe, W.} 1983,
\textit{MNRAS}, 202, 995

\bibitem[Kazantzidis, S. et al. (2004)]{Jafffe83}
{Kazantzidis, S., Kravtsov, A.V., Zentner, A.R., Allgood, B., Nagai, D., Moore, B.} 2004,
\textit{ApJ}, 611, 73

\bibitem[Mashchenko, S. et al. (2008)]{Mas08}
{Mashchenko, S., Wadsley, J., Couchman, H.M.P.} 2008,
\textit{Sci}, 319, 174

\bibitem[Navarro, J. F. et al. (1996)]{Nav96}
{Navarro J. F., Frenk C. S. \& White S.} 1996,
\textit{ApJ}, 462, 563

\bibitem[Navarro, J.F. et al. (2004)]{Nav04}
{Navarro, J.F., Hayashi, E., Power, C., Jenkins, A.R., Frenk, C.S., White, S., Springel, V., Stadel, J. \&  Quinn, T.R.} 2004,
\textit{MNRAS}, 349, 1039

\bibitem[Neto, A.F. et al. (2007)]{Neto07}
{Neto, A.F., Gao, L., Bett, P., Cole, S., Navarro, J.F., Frenk, C.S., White, S., Springel, V., \& Jenkins, A.} 2007,
\textit{MNRAS}, 381, 1450

\bibitem[Merrit, D. et al. (2006)]{Merrit06}
{Merritt, D., Graham A., Moore B., Diemand J. \& Terzic B.} 2006,
\textit{AJ}, 132, 2685

\bibitem[O\~norbe, J. et al. (2007)]{Ono07}
{O\~norbe, J., Domínguez-Tenreiro, R., Sáiz, A. \& Serna, A.} 2007,
\textit{AJ}, 376, 390

\bibitem[Prada, F. et al. (2006)]{Prada06}
{Prada, F., Klypin, A.A., Simonneau, E., Betancort-Rijo, J., Patiri, S., Gottler, S., Sanchez-Conde, M.A.} 2006,
\textit{ApJ}, 645, 1001

\bibitem[Read, J.I. \& Gilmore, G. (2005)]{Read05}
{Read,J.I. \& Gilmore,G.} 2005,
\textit{MNRAS}, 356, 107

\bibitem[Read, J.I. et al. (2006)]{Read06}
{Read, J.I., Pontzen, A.P., Viel, M.} 2006,
\textit{MNRAS}, 371, 885

\bibitem[Sellwood, J. A. \& McGaugh, S. S. (2005)]{Sell05}
{Sellwood, J.A., McGaugh, S.S.} 2005,
\textit{MNRAS}, 356, 107

\bibitem[Scannapieco, C. et al. (2005)]{Scan05}
{Scannapieco C., Tissera P.B., White S. \& Springel V.} 2005, 
\textit{MNRAS}, 364, 552

\bibitem[Scannapieco, C. et al. (2006)]{Scan06}
{Scannapieco, C., Tissera P.B., White S. \& Springel V.} 2006,
\textit{ApJ}, 634, 70

\bibitem[Scannapieco, C. et al. (2008)]{Scan08}
{Scannapieco, C. Tissera, P.B., White S. \& Springel V.} 2008,
\textit{MNRAS}, 389, 1137

\bibitem[Springel, V. et al. (2008)]{Sprin08}
{Springel, V., White, S., Frenk, C. S., Navarro, J.F., Jenkins, A., Vogelsberger, M., Wang, J., Ludlow, A., Helmi, A.} 2008,
\textit{Nature}, 456, 73

\bibitem[Springel, V. et al. (2008)]{Sprin08b}
{Springel, V., Wang, J., Vogelsberger, M., Ludlow, A., Jenkins, A., Helmi, A., Navarro, J. F., Frenk, C. S., White, S.} 2008b,
\textit{MNRAS}, 391, 1685

\bibitem[Tissera, P. B. \& Dominguez-Tenreiro, R. (1998)]{Ti98}
{Tissera, P.B. \& Dominguez-Tenreiro, R.} 1998,
\textit{MNRAS}, 297, 177

\bibitem[Young, P. (1980)]{Yo80}
{Young, P.} 1980,
\textit{ApJ}, 242, 1232

\end{thebibliography}
\end{document}